\documentclass[aps,prb,reprint,superscriptaddress]{revtex4-2}

\usepackage[utf8]{inputenc}
\usepackage[T1]{fontenc}
\usepackage{lmodern}
\usepackage{graphicx}
\usepackage{enumerate}
\usepackage[explicit]{titlesec}
\usepackage{blindtext,color}
\usepackage{hyperref}
\usepackage{amsmath}
\usepackage{amssymb}
\usepackage{amsthm}
\usepackage{bm}
\usepackage {tensor}
\usepackage{etoolbox}
\usepackage{enumitem}
\usepackage[english]{babel}
\usepackage{dsfont}
\usepackage{float}
\usepackage{slashed,braket}
\usepackage{stackengine}
\usepackage{mathrsfs}
\usepackage{verbatim}
\usepackage{array}
\usepackage{upgreek}
\usepackage{bbm}
\usepackage{subcaption}
\numberwithin{equation}{section}

\newcommand{\beq}{\begin{equation}}
\newcommand{\eeq}{\end{equation}}
\newcommand{\beqa}{\begin{eqnarray}}
\newcommand{\eeqa}{\end{eqnarray}}

\newcommand{\Tr}{\operatorname{Tr}}

\begin{document}

\title{Valley Hall viscosity in the integer quantum Hall phases of (2+1)D Dirac materials}

\author{M. Selch}
\affiliation{Bar-Ilan University, Tel Aviv, Israel}


\begin{abstract}
We calculate the valley-resolved Hall viscosity for Lorentz-invariant integer quantum Hall phases in Semenoff-semiconducting graphene-like systems at zero temperature. The Kubo formalism based discussion reported in Phys. Rev. B {\bfseries 100}, 115421 (2019) revealed the divergence of single valley viscous Hall contributions for this case with only a valley-summed Hall viscosity being finite and therefore well-defined. Our approach to the Hall viscosity calculation is based on an equivalent Green function formulation within Wigner-Weyl calculus. We find that the previously identified divergence may be regularized in a way compatible with both valley-summed Hall viscosity and (non)local Hall conductivity. Together with the local Hall conductivity and its first nonlocal correction, reported as well in Phys. Rev. B {\bfseries 100}, 115421 (2019), we extend the empirical relativistic Hoyos-Son formula to individual valleys. Both the original Hoyos-Son formula for Galilean invariant fluids and its relativistic extension to Dirac materials are found to be structurally identical for integer quantum Hall phases and expressible in terms of local electric and viscous Hall responses. In addition we evaluate the valley(-difference) Hall viscosity for biased Bernal bilayer graphene in the chiral fermion low energy approximation using the same regularization prescription. Prospects of measuring valley Hall viscosity in nonlocal transport for mono- and bilayer graphene- and group-VI TMD-based devices are discussed. 
\end{abstract}

\maketitle
\section{Introduction}
The field of valleytronics explores transport phenomena based on carriers with a valley charge in distinction to electric charge and spin. Valley-selective transport can be realized naturally in so-called Dirac materials which are characterized by the presence of energy band touching or nearly touching (avoided crossing) points with Dirac-like dispersion at (or close to) their Fermi points (surface). Each such point may be considered a valley in the electronic band structure whose electron or hole Fermi pockets carry electronic degrees of freedom relevant for low energy transport.\par
A paradigmatic 2D Dirac material is graphene \cite{castroneto2009the,goerbig2011electronic}. It is well-known that pristine graphene at charge neutrality is approximated by a Dirac semimetal with two physically distinct isoenergetic Fermi points located in its Brillouin zone (Fig. \ref{valleyhallfigure}, top right). The Dirac characteristics of graphene around charge neutrality originate from its lattice structure. Its unit cell consists of two atomic sites as part of a honeycomb lattice structure with the two sublattices A and B spanning a two dimensional  (pseudo)Dirac space (Fig. \ref{valleyhallfigure}, top left). Integrated into two dimensional heterostructures, placed on certain 3D substrates or exposed to external fields and distinct chemical environments, graphene is known to become gapped allowing for distinct massive topological phases. There are several well-known microscopic mechanisms by which graphene may acquire a sizable band gap. A staggered sublattice potential produces an asymmetry between the sublattices which breaks inversion symmetry and gives rise to a topologically trivial semiconducting phase, the Semenoff massive phase \cite{semenoff1984condensed}. An experimental realization is graphene aligned with hexagonal boron nitride (hBN). Another relevant class of Dirac materials are group-VI transition metal dichalcogenides (TMD's) (MX$_2$ with M=Mo,W and X=S,Se) which are naturally gapped into a Semenoff-massive phase in orbital rather than sublattice space and share symmetry characteristics with gapped graphene \cite{xiao2012coupled,cai2013magnetic}.\par
Notably electric valley Hall transport has been predicted \cite{xiao2007valley,song2014topological,song2016giant} and observed in several 2D Dirac materials such as graphene-based heterostructures including graphene on top of hexagonal boron nitride \cite{gorbachev2014detecting} and bilayer graphene \cite{shimazaki2015generation,yin2022tunable,sui2015gate} as well as transition metal dichalcogenides \cite{mak2014the} via nonlocal transport experiments. Signatures consistent with valley Hall currents in both monolayer and bilayer graphene devices with broken inversion symmetry are mediated via large valley-contrasting Berry curvature hot spots. 
These developments suggest that valley-dependent transport coefficients can provide direct probes of momentum-space geometry beyond conventional charge and spin transport.\par
A closely related geometric transport coefficient is the Hall viscosity, a non-dissipative viscosity that appears in systems with broken time-reversal symmetry. It induces transverse momentum transport via a viscous Lorentz force acting on shear flow. Unlike ordinary shear viscosity, Hall viscosity does not contribute to entropy production and is instead associated with parity-odd stress responses. It was originally studied in quantum Hall fluids where it arises geometrically as a consequence of Berry curvature similar to Hall conductivity \cite{avron1995viscosity,avron1998odd,read2009non,read2011hall,bradlyn2012kubo,abanov2014electromagnetic,gromov2014density,hoyos2012hall,cho2014geometry,hoyos2014hall}, These studies revealed (at least) two noteworthy properties: First from a topological point of view the Hall viscosity to density ratio is quantized in the presence of translational and rotational symmetry and proportional to the so-called Wen-Zee ``shift''. Second Hall viscosity is intimately related to Hall conductivity as established by the Hoyos-Son relation derived via Ward identities and effective field theory methods for Galilean invariant systems \cite{hoyos2012hall,bradlyn2012kubo} which was subsequently extended empirically to the case of Dirac materials \cite{sherafati2016hall,sherafati2019hall}. More recently (Hall) viscous flow attracted attention in hydrodynamic electron systems with the prospect of application to 2D Dirac materials \cite{alekseev2016negative, scaffidi2017hydrodynamic,delacretaz2017transport,pellegrino2017nonlocal}. A major experimental advance was achieved in \cite{berdyugin2019measuring} wherein the first measurement of Hall viscosity in graphene through nonlocal magnetotransport in the hydrodynamic regime has been reported. This work demonstrated that viscous electron flow in weak magnetic fields produces local electrochemical potential signatures opposite in sign to those of the classical Hall effect, thereby establishing Hall viscosity as an experimentally accessible electronic transport coefficient. This work was recently extended to graphene superlattices \cite{kim2025viscous}.\par
Despite the progress in both electric valley Hall transport and electron induced Hall viscosity, the interplay between valley degrees of freedom and viscous transverse stress responses remains largely unexplored. In multivalley Dirac systems with valley-contrasting Berry curvature due to broken inversion symmetry, one may naturally expect the emergence of a valley Hall viscosity (Fig. \ref{valleyhallfigure}, bottom) analogous to the valley Hall conductivity. Such a quantity would encode geometric information of valley-polarized fluids and provide a hydrodynamic probe of Berry curvature physics extending valley Hall physics from conductive to viscous phenomena. The concept of valley Hall viscosity for gapped Dirac materials in the integer quantum Hall regime has been claimed ill-defined in \cite{sherafati2019hall}.\par
In this work, we show that valley Hall viscosity in inversion and time reversal symmetry broken massive Dirac materials tuned to integer quantum Hall phases can be extracted from a linear response analysis by employing a proper regularization prescription. Within this prescription only a finite number of states in the vicinity of the chemical potential contributes to the valley Hall viscosity and at the same time reproduces the valley-summed Hall viscosity. Moreover our choice implies a valley-selective empirical Hoyos-Son formula for integer quantum Hall phases following \cite{sherafati2019hall}. We discuss valley(-difference) relative to (valley-summed) Hall viscosity in graphene-based systems and TMD's as well as prospects for its detection in hydrodynamic electron flow. The possibility of measuring the valley Hall viscosity necessarily implies the existence of a physical, well-defined, finite value of this transport coefficient. We provide details on Wigner-Weyl calculus, the calculation of (valley) Hall viscosity and the (valley-resolved, relativistic) Hoyos-Son formula in the Appendix.

\section{Field theory of Dirac fermions}
\begin{figure}[H]
\hspace{0.0cm}
\begin{subfigure}{0.30\linewidth}
\includegraphics[width=\linewidth]{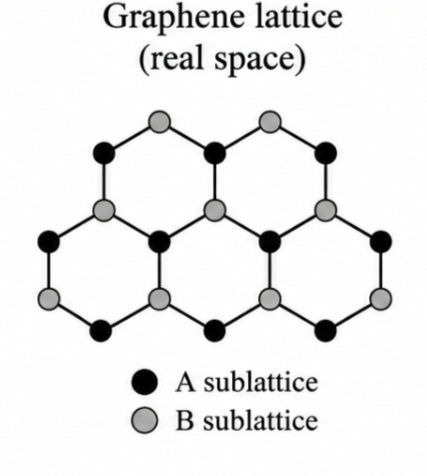}
\end{subfigure}
\begin{subfigure}{0.52\linewidth}
\hspace{-0.68cm}
\includegraphics[width=\linewidth]{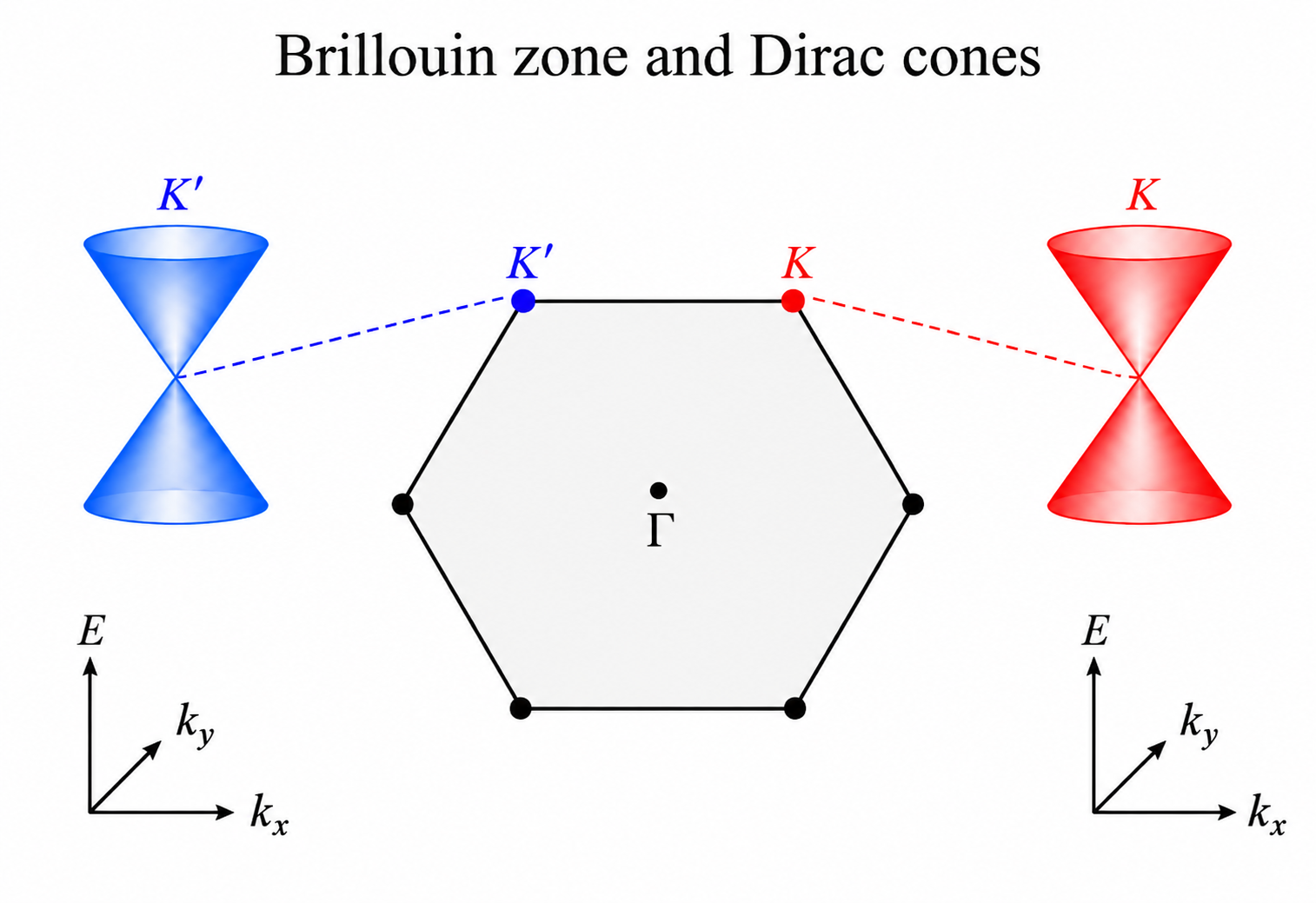}
\end{subfigure}
\begin{subfigure}{1.0\linewidth}
\includegraphics[width=\linewidth]{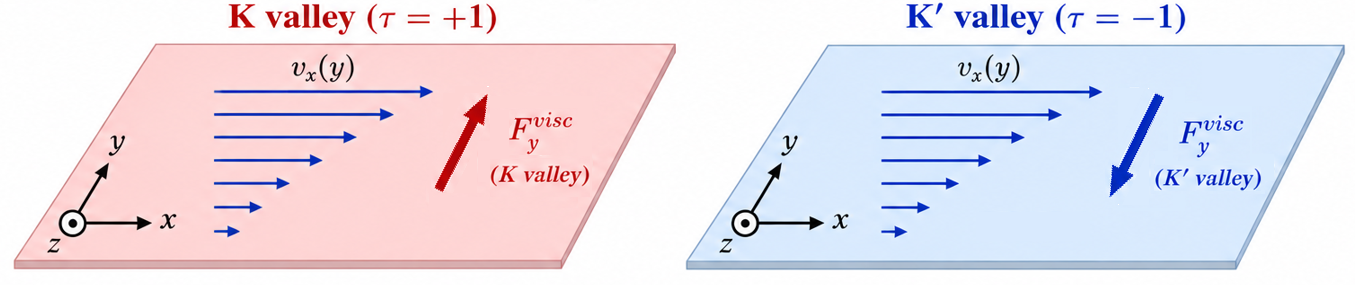}
\end{subfigure}
\caption{Valley Hall viscosity in graphene. Top left: Graphene's hexagonal lattice consisting of two interpenetrating triangular sublattices. Top right: Graphene's hexagonal Brillouin zone with Dirac cone dispersion at alternating K- and K'-points at the corners. At charge neutrality, the Fermi energy coincides with the cone points. Bottom: Antiparallel K- and K'-valley viscous Lorentz forces with valley index $\tau =\pm$ induced by shear flow in an external magnetic field (not shown) along the z-direction.}
\label{valleyhallfigure}
\end{figure}

We consider graphene and TMD's as representatives of (2+1)D Dirac materials which are described by a Dirac theory at low energies close to charge neutrality (Fig. \ref{valleyhallfigure}, top right). These may serve as a starting point to understanding Dirac materials more generally. Pristine graphene at charge neutrality (half-filling) is approximately a semimetal exhibiting two inequivalent band touching points in its Brillouin zone which define two valleys. A Dirac mass for graphene may be induced by placing it on a substrate such as hBN \cite{jung2017moire}, whereas Dirac fermions in group-VI TMD's are naturally massive (see table I in \cite{xiao2012coupled}). As a simplified model we consider an ideal Semenoff semiconductor parametrized by a constant and finite mass $m$ of opposite signs in the two valleys. We therefore regard moir\'e potentials for graphene on hBN and spin-orbit coupling for TMD's as negligible perturbations of our model.
In the vicinity of their would be Fermi points the low energy field theory action of non-interacting Semenoff semiconducting graphene or TMD's in Minkowski spacetime at chemical potential $\mu$ with mass parameter $m$ may be written in the Dirac form
\begin{align}
\nonumber S_M=&\int d^3xe\Psi^{\dagger}Q_M\Psi =\int d^3xe\Psi^{\dagger}(\omega -H_M)\Psi \\
=&\int d^3xe\bar{\Psi}(i\gamma_M^ae_a^{\mu}D_{\mu}-\gamma_M^0\mu +imv_F^2\tau^3)\Psi 
\label{minkowskidiracaction}
\end{align}
with Dirac spinors $\Psi$, $\Psi^{\dagger}$ and $\bar{\Psi}=\Psi^{\dagger}\gamma_M^0$, vielbein field $e_a^{\mu}=\bar{e}_a^{\mu}+\delta e_a^{\mu}$, its inverse determant $e=\det^{-1}(e_a^{\mu})$ and covariant derivative $D_{\mu}=\partial_{\mu}+iA_{\mu}+i\tau^3A^u_{\mu}+i(\tau^3)\delta A^{(u)}_{\mu}$. The Dirac spinors span sublattice/orbital ($\sigma$), valley ($\tau$) and spin spaces each of two dimensions. We may therefore introduce Pauli matrices subject to each of these spaces in general. The Dirac $\gamma$-matrices refer to sublattice/d-orbital space with the structure in valley and spin spaces being trivial for semimetallic graphene, while a Semenoff mass may be represented as a valley-odd parameter. A convenient representation of Minkowski spacetime Dirac $\gamma$-matrices is $\gamma_M^0=i\sigma^3$, $\gamma_M^1=\sigma^2$, $\gamma_M^2=-\sigma^1$ for $(-,+,+)$ signature which implies $\gamma_M^0\gamma_M^i=\sigma^i$,  $i=1,2$. The vierbein field is assumed to be trivial such that $\bar{e}_a^0=\frac{1}{\bar{e}}\delta_a^0$ and $\bar{e}_a^i=\frac{v_F}{\bar{e}}\delta_a^i$ with Fermi velocity $v_F$. The electromagnetic gauge field $A_{\mu}$ is induced by a constant background magnetic field supplemented by an emergent valley charged strain gauge field $A^u_{\mu}$ assumed to originate from external strain application such that it provides a constant background valley-axial pseudomagnetic field. Probe fields are denoted by $\delta e_a^{\mu}$ and $\delta A^{(u)}_{\mu}$, respectively. Electromagnetic and strain gauge fields are naturally associated with valley-even and valley-odd charge transport currents by variation of the (effective) action with respect to the corresponding probe fields. \par
Within the Dirac fermion approximation, strain may be encoded geometrically implying emergent vielbein field modifications (and therefore emergent metric modifications \cite{leyva2015generalizing,volovik2015emergent,zubkov2015emergent,juan2013gauge}) which we do not consider in this work. 
In-plane strain as parametrized by a two dimensional displacement field $u_i(x)$ $(i=1,2)$ with corresponding strain field $u_{ij}=\frac{1}{2}(\partial_iu_j+\partial_ju_i)$ implies an
emergent strain gauge field which takes the explicit form
\begin{align}
A_i^{u}=e^uK_{ijk}\epsilon_{kl} u_{jl}.
\label{straingaugefield}
\end{align}
for graphene. The Levi-Civita symbol is defined by $\epsilon_{12}=-\epsilon_{21}=1$. The coefficients $K_{ijk}$ with non-vanishing components given by $K_{111}=-K_{122}=-K_{212}=-K_{221}=1$ contain information about the graphene lattice. The strain charge modulus $e^u=\beta /(2a)$ comprises the fundamental lattice constant $a$ as well as the Grüneisen parameter $\beta$ of the system. A constant background magnetic field as a time reversal symmetry breaking source for Hall viscosity stands in analogy to a constant background strain-induced pseudomagnetic field sourcing the time reversal even valley Hall viscosity.\par
The non-relativistic limit of Eq. (\ref{minkowskidiracaction}), understood as a standard non-relativistic approximation, with non-relativistic chemical potential $\bar{\mu}=\mu -m$ receives the form
\begin{align}
\nonumber &S_M= \int d^3 x \sqrt{-g}\Psi^{\dagger}Q_M \Psi =\int d^3 x\sqrt{-g} \Psi^{\dagger}(\omega -H_M) \Psi\\
&=\int d^3 x\sqrt{-g} \Psi^{\dagger}(iD_0+\bar{\mu} +\frac{1}{2m}g^{ij}D_iD_j+\frac{1}{2m}\tau^3eB)\Psi \label{SMink}
\end{align}
with inverse metric tensor $g^{\mu\nu}=e_a^{\mu}\eta^{ab}e_b^{\nu}$, inverse Minkowski metric $\eta^{ab}$ and $\sqrt{-g}=e$. 
The magnetic field is seen to induce a Zeeman term in valley space.\par
The Euclidean space formulation is obtained by a Wick rotation with time coordinate $t$ replaced by an imaginary time coordinate $\tau =it$. The Euclidean space Dirac $\gamma$-matrices are related to those in Minkowski spacetime via $\gamma_E^0=i\gamma_M^0$, $\gamma_E^i=\gamma_M^i$. The Euclidean space action is given by
\begin{align}
\nonumber -S_E=&\int d^3xe\Psi^{\dagger}Q_E\Psi =\int d^3xe\Psi^{\dagger}(i\omega -H_E)\Psi \\
=&\int d^3xe\bar{\Psi}(\gamma_E^ae_a^{\mu}D_{\mu}+\gamma_E^0\mu +mv_F^2\tau^3)\Psi .\label{SED}
\end{align}
The conjugate spinor is defined by $\bar{\Psi}=\Psi^{\dagger}\gamma_E^0$. In the non-relativistic limit this amounts to the action
\begin{align}
\nonumber &-S_E=\int d^3 x\sqrt{g} \Psi^{\dagger} Q_E\Psi =\int d^3 x\sqrt{g} \Psi^{\dagger}(i\omega -H_E) \Psi\\
&=\int d^3 x\sqrt{g} \Psi^{\dagger}(-D_0+\bar{\mu} +\frac{1}{2m}g^{ij}D_iD_j+\frac{1}{2m}\tau^3eB)\Psi . \label{SEucl}
\end{align}
While the spin degrees of freedom enter equally, we find a Pauli action in valley space with a valley Zeeman term. This implies a spin degenerate fermion theory with two valley species whose Landau levels are shifted relative to each other by one in an external background magnetic field.\par 
We may add Coulomb interactions to the actions of Eqs. (\ref{SED}) and (\ref{SEucl}) by replacing $\mu$/$\bar{\mu}$ by $\mu$/$\bar{\mu}-\lambda(x)$ and adding the term 
\begin{align}
S[\lambda ]=&\frac{1}{2} \int dtd^3\bold{x} d^3\bold{x}^{\prime} \lambda(t,\bold{x}) V^{-1}(\bold{x},\bold{x}^{\prime })\lambda(t,\bold{x}^\prime).
\end{align}
Such a description is obtained naturally from a continuum limit derived from lattice regularized Dirac materials in the presence of Coulomb interactions. The field $\lambda$ is the dynamical density fluctuation introduced by a Hubbard-Stratonovich transformation. It couples to fermions by a Yukawa term and mediates Coulomb interactions whose strength is determined by the Coulomb potential $V$ with inverse $V^{-1}=\frac{1}{e^2}\Delta^{(3)}$. The latter represents the kinetic energy operator of the Hubbard-Stratonovich field. The operator $\Delta^{(3)}$ is the three dimensional Laplacian and $e$ is the electric charge.\par
We employ(ed) units such that $\hbar =c=1$ with $\hbar$ the reduced Planck constant and the velocity of light $c$. We will suppress the electric charge $e\to 1$ and for notational convenience furthermore the Fermi velocity $v_F$ for most of the discussion.

\section{Hall viscosity in linear response theory within Wigner-Weyl calculus}
We will extract an expression for the valley Hall viscosity in integer quantum Hall phases and evaluate it explicitly. We begin with the definition of the stress tensor
\begin{align}
T_i^a(x)=-\frac{\delta\log Z}{\delta e_a^i(x)}=-\frac{1}{Z}\frac{\delta Z}{\delta e_a^i(x)}
\end{align}
with Euclidean partition function $Z$. In the linear response approximation the Hall viscosity may be extracted from the averaged stress tensor $\bar{T}_i^a=\frac{1}{\beta_TA}\int d^3xT_i^a(x)$ according to (see Appendix \ref{appendixa})
\begin{align}
\eta_H=&\frac{1}{4}\epsilon_{ab}\delta^{ij}\frac{\partial \bar{T}_i^a}{\partial (\partial_0e_b^j)}
=\frac{1}{2\pi}\mathcal{N}_{\eta}B
\label{valleyhallviscosity}
\end{align}
with background magnetic field $B$, inverse temperature $\beta_T$ and sample area $A$. $\delta^{ij}$ is the Kronecker-Delta symbol. The calligraphic coefficient $\mathcal{N}_{\eta}$ entering Eq. (\ref{valleyhallviscosity}) is expressed in terms of Weyl-symbols of operators, more specifically the fermion bilinear operator $\hat{Q}$, its inverse $\hat{G}$, which is the fermion propagator, and the covariant derivative. Weyl-symbols of operators are functions on phase space which are in one-to-one correspondence with their respective operators. They reduce to momentum space representations of operators in the homogeneous limit (here for $B\to 0$). A product of operators becomes a Moyal star product of Weyl-symbols.
For operators $\hat{A}$ and $\hat{B}$ we have $(\hat{A}\hat{B})_W=A_W\star B_W$ with 
\begin{align}
\star = {\rm exp} \Big(\frac{i}{2} \left( (\overleftarrow{\partial_x})^{i}(\overrightarrow{\partial_k})_{i}-(\overleftarrow{\partial_k})_{i}(\overrightarrow{\partial_x})^{i}\right) \Big).
\end{align}
For the fermion bilinear and its inverse this implies the Groenewold equation $Q_W\star G_W=G_W\star Q_W=1$ on phase space. We consider the simplification of phase space dependence of Weyl operators being reduced to $i(D_W)_{\mu}=k_{\mu}-A_{\mu}(x)\equiv (\pi_W)_{\mu}$, which is the kinetic momentum, meaning magnetic translation invariance. In the presence of a background magnetic field not all momenta can be made good quantum numbers. The Moyal star product is associative but not commutative which is the phase space manifestation of operator ordering. Together with the reduced phase space dependence of Weyl symbols this is expressed as
\begin{align}
\star =
\exp\Big(-\frac{i}{2}B\epsilon_{ij}\overset{\leftarrow}{\partial}_{(\pi_W)_i}\overset{\rightarrow}{\partial}_{(\pi_W)_j}\Big).
\end{align}
The coefficient $\mathcal{N}_{\eta}$ written in Weyl-symbol and differential form notation with $\omega =p_3$ reads (see Appendix \ref{appendixa})
\begin{align}
\nonumber \mathcal{N}_{\eta}=&\frac{1}{16\pi^2\beta_T A}\int d^3xd^3k\frac{\epsilon^{ij}\delta^{lm}}{B}\Tr\Big[\frac{\partial Q_W}{\partial k_i}\star (D_W)_l \\
\nonumber &\star G_W\star\frac{\partial Q_W}{\partial k_j}\star (D_W)_m\star G_W\star \frac{\partial Q_W}{\partial k_3}\star G_W\Big]\\
\nonumber =&\frac{1}{48\pi^2\beta_T AB}\int d\tau\wedge dx^1\wedge dx^2\wedge\Tr\Big[dQ_W\star (\bold{D}_W)^T\\
&\star G_W\wedge\star dQ_W\star (\bold{D}_W)\star G_W\wedge\star dQ_W\star G_W\Big].\label{topologicalhallvis}
\end{align}
The coefficient  $\mathcal{N}_{\eta}$ represents the Hall viscosity topological invariant faithfully for Lorentz (and Galilean) invariant systems and whose Green function expression has been introduced in \cite{selch2026nonrenormalization} for the first time. In terms of the more common Landau level filling factor $\nu$, the Euler characteristic $\chi$, the charge carrier number $N$ and the number of flux quanta $N_{\Phi}=\frac{BA}{\Phi_0}$ with flux quantum $\Phi_0=2\pi$ the following relations are valid
\begin{align}
N=\nu N_{\Phi}+\mathcal{S}\frac{\chi}{2}\,\,\Rightarrow\,\,\mathcal{N}_{\eta}=\frac{1}{4}\nu \mathcal{S}.
\end{align}
The first relation defines the Wen-Zee shift $\mathcal{S}$. On a spatial torus $\chi =0$. Note the relation $\nu_D=g_{sv}\Big(p-\frac{1}{2}\Big)$ for Dirac fermions in graphene with valley and spin degeneracy factor $g_{sv}=4$ and $p-1$ filled Landau levels above the zero Landau level. 

\section{Explicit calculation of the valley(-resolved) Hall viscosity in Dirac materials}
The K- and K'-valley Hall viscosities may be understood as viscous Lorentz-like forces acting on fluid elements in a shear flow possibly induced by an external magnetic field (Fig. \ref{valleyhallfigure}, bottom). The individual valley Hall viscosities may be obtained as $\eta_H^K=\eta_H(-m)$ and $\eta_H^{K^{\prime}}=\eta_H(m)$, respectively, with the zero Landau level energies in the two valleys ordered according to $E_0^K<0<E_0^{K^{\prime}}$. We define the Hall viscosity as the valley-summed Hall viscosity $\eta^s_H=\eta_H^K+\eta_H^{K^{\prime}}$ and the valley Hall viscosity as the valley-difference Hall viscosity $\eta^d_H=\eta_H^{K^{\prime}}-\eta_H^{K}$. While the Hall viscosity is odd under time-reversal symmetry, the valley Hall viscosity is even, in analogy to the corresponding Hall conductivities, and therefore odd respectively even under directional flip of the magnetic field $B\to -B$. They are furthermore odd respectively even under particle-hole inversion. Semimetallic graphene at $\mu =0$ is particle-hole symmetric. Therefore its Hall viscosity vanishes. Semenoff massive graphene or TMD's have a particle-hole symmetric eigenvalue spectrum up to the unpaired zero Landau level within the considered simplified model. The massive Dirac fermions have opposite masses in the two valleys. The zero mode is shifted upward from zero in one valley ($K^{\prime}$) but downward from zero in the other valley ($K$) as soon as a Semenoff mass $m\tau^3$ is introduced. In the gapped phase we have a symmetry under particle-hole conjugation and simultaneous valley exchange. This implies vanishing Hall viscosity within the gap $-|m|<\mu <|m|$, whereas the valley Hall viscosity may be nonzero. We may consider electron ($\mu >|m|$) or hole ($\mu <-|m|$) doping relative to this gap and restrict ourselves to electron or n-doping due to symmetry.
We evaluate the Hall viscosity coefficient $\mathcal{N}_{\eta}$ in the limit of vanishing temperature using symbolic python.\par 
We start with a discussion of the valley-summed Hall viscosity which may be directly compared with the results in \cite{sherafati2016hall,sherafati2019hall}. For $|E_{p-1}|<\mu <E_{p}$ with $p\geq 1$ we obtain $\mathcal{N}^s_{\eta}=\frac{1}{4}(p^2+(p-1)^2)$ per spin degree of freedom. In comparison to \cite{sherafati2016hall,sherafati2019hall} we find a Hall viscosity smaller by a factor of four. This follows directly from their definition of the stress tensor which we consider to be twice as large as it should be. Proper normalization in our case is confirmed by a consideration of the corresponding non-relativistic calculations for which Eq. (\ref{topologicalhallvis}) is valid as well. In the non-relativistic limit the electrons of the two valleys have a relative Landau level occupation of one due to the valley Zeeman term in Eq. (\ref{SEucl}). This implies $\mathcal{N}_{\eta}^K=\frac{1}{4}(p-1)^2$ and $\mathcal{N}^{K^{\prime}}_{\eta}=\frac{1}{4}p^2$ such that $\mathcal{N}^s_{\eta}=\frac{1}{4}(p^2+(p-1)^2)$ per spin degree of freedom.\par
\begin{figure}[H]
\includegraphics[width=\linewidth]{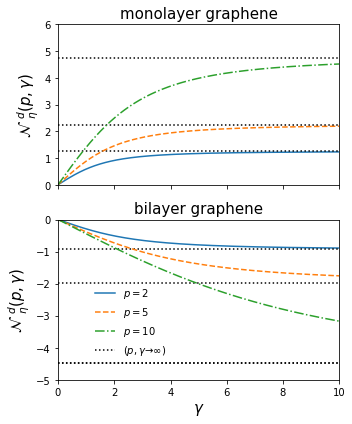}
\caption{Valley Hall viscosity coefficient $\mathcal{N}_{\eta}^d(p,\gamma)$ as a function of representative Landau level indices $p$ and inversion symmetry breaking parameter $\gamma$ for monolayer graphene (on hBN or TMD's) and bilayer graphene in the low energy chiral fermion approximation. Asymptotic values for $\gamma\to\infty$ are also shown.}
\label{valleyhallcoefficientplot}
\end{figure}

We now turn to the valley-resolved Hall viscosities. We derived a formula (see Appendix \ref{appendixa}) which allows for an algebraically simple determination of the Hall viscosity and its individual valley contributions.
The valley-resolved Hall viscosities of Dirac materials are divergent and so is the valley Hall viscosity (previously found in \cite{sherafati2019hall}), even though they may be determined experimentally and therefore require a natural regularization prescription. We propose such a prescription, elaborated from a calculation-based point of view in Appendix \ref{appendixb}, which is natural in the following sense: (i) It reproduces the (valley-summed) Hall viscosity, (ii) it coincides with the (valley) Hall viscosity in the non-relativistic limit and (iii) it is compatible with nonlocal Hall viscosity establishing a valley-resolved Hoyos-Son relation. The (valley) Hall viscosity may be expressed as a sum of terms involving an occupied and an unoccupied energy eigenstate within linear response theory. We may distinguish positive and negative energy eigenstate sectors relative to the zero Landau level in each valley. Sherafati and Vignale \cite{sherafati2019hall} distinguish three types of contributions in the sum: (i) terms where both empty and occupied levels belong to the positive sector, (ii) terms where the occupied level belongs to the negative sector, whereas the unoccupied level belongs to the positive sector and (iii) terms with contributions from the occupied zero Landau level with unoccupied level in the positive sector. Of these three cases only the first one appears in the non-relativistic limit. We consider terms of the first and third type and divide terms of the second type further: (ii1) terms for which the occupied level has a particle-hole conjugate partner contribution to type (i) and (ii2) all other terms. While the number of terms in (ii1) is finite, it is infinite in (ii2). We consider this splitting to be very natural and our prescription implies to drop all terms of type (ii2). 
We then find the so renormalized valley-resolved Hall viscosities to be finite, as only a two Landau level neighborhood of the chemical potential modulo particle-hole partner states is relevant to evaluate them. This may be considered a ``minimally relativistic'' prescription extending the sum from terms of type (i) to their relativistic particle-hole partner states modulo contributions involving the zero Landau level. The valley-resolved Hall viscosities are therefore determined locally in Hilbert space by a finite and small number of eigenfunctions and their energy eigenvalues. To the best of our knowledge this prescription has not been suggested before. It implies especially that the individual valley Hall viscosities vanish within the mass gap $-|m|<\mu <|m|$. With $\gamma =\frac{mv_F}{\sqrt{2eB}}$ we obtain for electron doping 
\begin{align}
\nonumber \mathcal{N}^{\zeta}_{\eta}=&\frac{1}{8}\Big[p^2+(p-1)^2\\
&-s(\zeta )\Big(\frac{(p-1)\gamma}{\sqrt{p+\gamma^2}}+\frac{p\gamma}{\sqrt{p+1+\gamma^2}}\Big)\Big]
\end{align}
per spin degree of freedom with $s(\zeta )=1$ for $\zeta =K$ and $s(\zeta )=-1$ for $\zeta =K^{\prime}$. The valley Hall viscosity has a coefficient $\mathcal{N}^d_{\eta}$ of the form
\begin{align}
\mathcal{N}^d_{\eta}=\frac{\gamma}{4}\Big(\frac{p-1}{\sqrt{p+\gamma^2}}+\frac{p}{\sqrt{p+1+\gamma^2}}\Big)
\label{valleyhallvis}
\end{align}
per spin degree of freedom (Fig. \ref{valleyhallcoefficientplot}, top). The valley Hall viscosity is odd under simultaneous valley and particle-hole exchange. In the non-relativistic limit $\gamma \gg 1$ such that $\mathcal{N}^d_{\eta}=\frac{2p-1}{4}=\frac{1}{4}(p^2-(p-1)^2)$. This is one of our main results. To sum up the deep Dirac sea comprising piece in the sum representing valley-resolved Hall viscosities must be removed to obtain the transport Hall viscosities of the individual valleys. The cut-off of the infinite sum for occupied states is made at an eigenenergy beyond the mirror image of the chemical potential in the symmetric Dirac cone spectrum. Confidence in our result can be gained by comparison with the local and nonlocal valley Hall conductivities of massive Dirac fermions computed in \cite{sherafati2019hall} which is discussed shortly and supplemented by  the finite frequency case in Appendix~\ref{appendixc}. The renormalized valley-resolved Hall viscosities may then be equivalently understood as the unique valley contributions to the nonlocal valley-resolved Hall conductivity naturally associated to Hall viscosity.

\section{Valley-resolved Hoyos-Son formula}
The Hoyos-Son formula \cite{hoyos2012hall} establishes a connection between the first nonlocal correction of the Hall conductivity and the Hall viscosity. This relation has been proven rigorously for Galilean invariant fluids but not in the context of Dirac materials exhibiting Lorentz invariance. In the latter case an analogous relation has been established empirically by direct computation. To achieve this on the valley-resolved level we employ results from \cite{sherafati2019hall}, who evaluated the required nonlocal Hall conductivity, and combine them with our independent calculation. In the limit of large filling factor $p\gg 1$ we will find results that may be physically interpreted in the way originally proposed by Hoyos and Son \cite{hoyos2012hall}.
We begin with the expression for the local Hall conductivity $\sigma_H=\lim\limits_{k\to 0}\sigma_H(k)$ with momentum $k$ which takes the well-known form $\sigma_H=\frac{1}{2\pi}\mathcal{N}_{\sigma}$ with Chern number $\mathcal{N}_{\sigma}$. The topological coefficient $\mathcal{N}_{\sigma}$ written in Weyl-symbol notation with $\omega =k_3$ reads
\begin{align}
\nonumber \mathcal{N}_{\sigma}=&\frac{1}{24\pi^2\beta_T A}\int d^3xd^3k\epsilon_{\mu\nu\rho}\Tr\Big[\frac{\partial Q_W}{\partial k_{\mu}} \\
&\star G_W\star\frac{\partial Q_W}{\partial k_{\nu}}\star G_W\star\frac{\partial Q_W}{\partial k_{\rho}}\star G_W\Big].\label{topologicalhallcon}
\end{align}
The valley-resolved local Hall conductivities following \cite{sherafati2019hall} are given by
\begin{align}
&\sigma_H^{\zeta}=\frac{e^2}{2h}\Big(2p-1-s(\zeta )\frac{\gamma}{p+\gamma^2}\Big).
\end{align}
We suggest a relativistically generalized Hoyos-Son formula as advertised in \cite{sherafati2016hall,sherafati2019hall} with additional valley resolution. For integer quantum Hall phases we find the new result
\begin{align}
\sigma^{\zeta}_H(k)=\sigma^{\zeta}_H+(kl_B)^2\Big[\frac{e^2}{B}\eta^{\zeta}_H-\Big(p-\frac{1}{2}\Big)\sigma_H^{\zeta}\Big].
\label{valleyresolvedhoyosson}
\end{align}
We employed the expression for the charge carrier density $n=g_{sv}(p-\frac{1}{2})\frac{B}{2\pi}$ in deriving Eq. (\ref{valleyresolvedhoyosson}) and indicate valleys by $\zeta =K,K^{\prime}$. This form is found to be identical for integer quantum Hall phases of Galilean invariant fluids \cite{hoyos2012hall}, though the physical interpretation of the relativistic result and for the general case including fractional filling fractions remains to be found. 
We may consider the valley-summed $\sigma_H^s(k)=\sigma_H^K(k)+\sigma_H^{K^{\prime}}(k)$ and valley-difference $\sigma_H^d(k)=\sigma_H^{K^{\prime}}(k)-\sigma_H^{K}(k)$ (non)local Hall conductivities. For $p\gg 1$ and consulting \cite{sherafati2019hall} we find up to $O(k^4)$ for the valley- and spin-summed Hoyos-Son formula for graphene
\begin{align}
\frac{\sigma_H^s(k)}{g_{sv}\frac{e^2}{h}}=\Big(p-\frac{1}{2}\Big)\Big[1+(kl_B)^2\Big(\frac{\eta^s_H}{n}-(p-\frac{1}{2})\Big)\Big]
\end{align}
for the nonlocal Hall conductivity in the massless case and as well for the massive case so long as $p\gg max(1,\gamma^2)$. In the latter limit we recover the physical interpretation of the constituents of the Hoyos-Son formula as those for Galilean invariant fluids. The second contribution to the nonlocal piece is proportional to the inverse internal compressibility (evaluated at fixed filling fraction), while the inverse compressibility (evaluated at fixed charge carrier density) diverges which is why integer (or fractional) Hall fluids are termed incompressible. Interestingly the internal compressibility for Hall fluids derived for Dirac materials is even in $\gamma$ (which is straightforward to see) and does therefore not enter in valley-difference formulae. The appearance of nonlocal contributions odd in $\gamma$ beyond the valley-difference Hall viscosity shows the breakdown of the correspondence of terms compared to Galilean invariant fluids.\par
The local valley Hall conductivity reads $\sigma^d_H=\frac{1}{2\pi}\mathcal{N}_{\sigma}^d$ with
\begin{align}
\mathcal{N}_{\sigma}^d=\mathcal{N}_{\sigma}^{K^{\prime}}-\mathcal{N}_{\sigma}^{K}=\frac{\gamma}{\sqrt{p+\gamma^2}}
\end{align}
per spin degree of freedom. For $p\gg 1$ the ratios $\mathcal{N}^d_{\sigma}/\mathcal{N}^s_{\sigma}$, $\mathcal{N}^d_{\eta}/\mathcal{N}^s_{\eta}\sim p^{-\frac{3}{2}}$ therefore obey the same scaling behavior. Notice the omission of a factor of four in our Hoyos-Son relations relative to \cite{sherafati2019hall} which we alluded to before and that the valley-resolved Hall conductivities and Hall viscosities explicitly depend on the Semenoff mass.

\section{Topological robustness}
The Hall conductivity $\sigma_H=\frac{1}{2\pi}\mathcal{N}_{\sigma}$ comes with the coefficient $\mathcal{N}_{\sigma}$, shown in Green function representation in Eq. (\ref{topologicalhallcon}), which represents the first Chern number and is a topological invariant.\par
The coefficient $\mathcal{N}_{\eta}$ entering the Hall viscosity is only topological under more restrictive conditions. Rotational invariance implies the general functional form $Q_W=Q_W(\omega ,\gamma^ae_a^i(D_W)_i)$ for Dirac fermions which, together with magnetic translation and Lorentz boost invariance, leads to the representation of $\mathcal{N}_{\eta}$ as given in Eq. (\ref{topologicalhallvis}). In analogy to the topological expression for the Hall conductivity we find a differential form representation for the integrand. 
Only the leading contribution of the valley-summed Hall viscosity in powers of the filling factor, expressed via Green functions in $\mathcal{N}_{\eta}$, has been shown to be topological under variations so far \cite{selch2026nonrenormalization}. Due to the assumption of Lorentz invariance the topology of the Green function expression for $\mathcal{N}_{\eta}$ follows from the well-known quantization of the Hall viscosity over particle density ratio under even weaker symmetry assumptions.
We expect the Hall viscosity to be protected against perturbative Coulomb interactions in analogy to the case of Galilean invariant fluids. This is motivated by the observation that the sum of the two valley contributions yields a contribution identical to that of its non-relativistic limit. As already mentioned, in this limit a sum of two independent (2+1)D Galilean invariant electron gases emerges whose Landau level spectra are relatively shifted by one unit due to the relative valley Zeeman splitting. In other words the two independent (2+1)D electron gases have opposite (valley-odd) orbital magnetic moments of the right magnitude for this coincidence to happen \cite{cai2013magnetic}.\par
The valley Hall viscosity is not topologically protected, as it depends explicitly on the Dirac mass. In the non-relativistic limit, however, the same expectation as for the Hall viscosity applies. This follows straightforwardly, because not only the sum but also the difference of two topological terms is again topological.

\section{Valley Hall viscosity of biased Bernal bilayer graphene}
We additionally consider Bernal bilayer graphene \cite{mccann2013the} in view of its theoretical and potential experimental relevance. The action of Bernal bilayer graphene is naturally constructed from two copies of monolayer graphene subjected to interlayer tunneling. Sublattice, valley and spin indices are supplemented by a layer index. We retain only the largest interlayer tunneling parameter $\gamma_1\approx 0.38$eV. The analog of inversion symmetry breaking compared to the case of monolayer graphene is achieved by applying a bias voltage between the two layers which induces a band gap $2\Delta =2mv_F^2$. We again set $v_F=1$. In the low bias and low energy limit $v_Fk\ll \Delta\ll \gamma_1$ with wave vector modulus $k$ Bernal bilayer graphene may be described by chiral fermions in a two component sublattice-layer subspace ($\sigma$). The Euclidean space action is given by \cite{mccann2013the}
\begin{align}
\nonumber -&S^{\zeta}_E=\int d^3xe\Psi^{\dagger}Q^{\zeta}_E\Psi =\int d^3xe\Psi^{\dagger}(i\omega -H^{\zeta}_E)\Psi \\
\nonumber =&\int d^3xe\Psi^{\dagger}\Big(-D_0+\mu +\frac{1}{\gamma_1}(e_x^iD_ie_x^jD_j-e_y^iD_ie_y^jD_j)\sigma^x\\
&+\frac{s(\zeta )}{\gamma_1}\{e_x^iD_i,e_y^jD_j\}\sigma^y+m\sigma^z\Big)\Psi .
\label{actionbilayergraphene}
\end{align}
Curly brackets denote the anti-commutator. Notice that while the action of Bernal bilayer graphene is in general Lorentz invariant, the low energy chiral fermion approximation breaks both Galilean and Lorentz symmetries. Therefore Eq. (\ref{topologicalhallvis}) represents the Hall viscosity coefficient faithfully only to leading order in the filling factor. We still choose to stick with the low energy theory, as the Hall viscosity calculation (or more precisely its leading contribution) is algebraically much less involved and yet seems to capture the valley Hall viscosity exactly. The simplicity of the action in Eq. (\ref{actionbilayergraphene}) allows for a straightforward and analogous calculation of the (valley) Hall viscosity coefficient of biased Bernal bilayer graphene as compared to monolayer graphene in the assumed regime of validity. We again employ symbolic python for our calculations.\par
Leaving some detail to Appendices \ref{appendixd} and \ref{appendixe} the Hall viscosity coefficient including degeneracies is given by
\begin{align}
\mathcal{N}_{\eta}^s(p)\overset{p\gg 1}{\approx}\frac{p^2}{2}
\end{align}
per spin degree of freedom in the limit of large filling factor which coincides with previous calculations \cite{hsiao2021timereversal}. 
Subleading terms should not be considered for this case. In order to parallel the discussion for Dirac materials we define the dimensionless parameter $\tilde{\gamma}=\frac{m\gamma_1}{2eB}$. The valley Hall viscosity coefficient evaluates to ($p\geq 3$)
\begin{align}
\nonumber \mathcal{N}_{\eta}^d=&-\frac{\tilde{\gamma}}{2}\Bigg[\frac{(p-2)(p-1)}{(2p-3)\sqrt{p(p-1)+\tilde{\gamma}^2}}\\
\nonumber &+\frac{(p-1)p}{(2(p+1)-3)\sqrt{(p+1)p+\tilde{\gamma}^2}}\Bigg]\\
\overset{p\gg 1}{\approx}&-\frac{\tilde{\gamma}p}{2\sqrt{p^2+\tilde{\gamma}^2}}.
\label{valleydiffbilayer}
\end{align}
per spin degree of freedom (Fig. \ref{valleyhallcoefficientplot}, bottom). The calculation of the coefficient in Eq. (\ref{valleydiffbilayer}) is another new result. Both by the structural similarity of the formula as compared to the case of Dirac fermions presented in Eq. (\ref{valleyhallvis}) and the proper scaling with magnetic field at constant carrier number density at large filling factor $\mathcal{N}_{\eta}^d\sim \frac{1}{B}$ are strong indicators in favor of our claim of exact representation. This is further undermined in Appendix \ref{appendixe}. To the best of our knowledge the corresponding local and nonlocal valley Hall conductivities have not yet been calculated via the Kubo (or an equivalent) formalism. We still expect the structural form of the Hoyos-Son formula as applied to non-dissipative valley electric and viscous transport to be identical to that found for Lorentz and Galilean invariant fluids, since this has been confirmed by explicit calculation in the absence of an interlayer bias \cite{hsiao2021timereversal}.

\section{Experimental prospects of measuring the valley Hall viscosity in hydrodynamic electron flow}
The electron Hall viscosity has been measured indirectly for the first time via nonlocal transport experiments in \cite{berdyugin2019measuring} for electron hydrodynamic flow within a graphene-based sample. More recently successful analogous experiments on Hall viscosity determination in graphene superlattices have been reported \cite{kim2025viscous}.
The number of occupied Landau levels $p$ relative to the zero Landau level below a fixed chemical potential scales as $p\sim \frac{1}{B}$. This implies that $\sigma^s_H,\,\eta^s_H\sim \frac{1}{B}$. A naive extrapolation of this scaling law beyond the achievable experimental resolution of Hall plateaus to magnetic fields of around $40$mT as used in \cite{berdyugin2019measuring, kim2025viscous} implies in accordance with Table I. in \cite{mechelen2019viscous} that indeed the valley-summed Hall viscosity of electrons has been extracted successfully within nonlocal transport.\par
Note that the hydrodynamic regime for collective viscous electron flow requires finite temperatures as well as large enough charge carrier densities. The extraction of valley Hall conductivity and valley Hall viscosity in the presence of a background magnetic field favors large gap sizes or effective masses.\par

The measurement of valley Hall viscosity via nonlocal transport for hydrodynamic electron flow is more difficult than that of Hall viscosity due to significantly smaller signal strength. Consider perfectly valley-polarized currents in the hydrodynamic regime ($p\gg 1$). A nonlocal (nl) resistance measurement with vicinity resistance
\begin{align}
R_A(B)=\frac{1}{2}(R_{nl}(B)-R_{nl}(-B)),\,\,\,\,R_{nl}=\frac{V_{nl}}{I_{nl}}
\end{align}
implies a relative signal strength of valley Hall viscosity over Hall viscosity of
\begin{align}
\Delta=&\Big\rvert\frac{R_A^{K^{\prime}}(B)-R_A^K(B)}{R_A^{K^{\prime}}(B)+R_A^K(B)}\Big\rvert=\Big\rvert\frac{\eta_H^{K^{\prime}}-\eta_H^K}{\eta_H^{K^{\prime}}+\eta_H^K}\Big\rvert =\Big\rvert \frac{\mathcal{N}^d_{\eta}}{\mathcal{N}^s_{\eta}}\Big\rvert .
\end{align}
The functional form of the vicinity resistance $R_A$ is provided explicitly in \cite{berdyugin2019measuring} from which that of $\Delta$ may be obtained straightforwardly. We find
\begin{align}
\Delta\sim\frac{|\gamma |}{p^{\frac{3}{2}}}=|\gamma |\Big(\frac{g_{sv}B}{2\pi n}\Big)^{\frac{3}{2}}\,\Big/ \, \sim\frac{|\tilde{\gamma} |}{p^2}=|\tilde{\gamma} |\Big(\frac{g_{svl}B}{2\pi n}\Big)^2
\end{align}
for Dirac materials (left) and biased Bernal bilayer graphene (right). Note that the hydrodynamic regime requires relatively small magnetic fields but relatively large charge carrier densities both of which disfavor large $\Delta$. With representative values for a magnetic field of $40$mT, a charge carrier density of $\frac{1}{2}\cdot 10^{12}$cm$^{-2}$ and $g_{svl}=2g_{sv}=8$ we find 
\begin{align}
\Delta\approx 8\cdot \Big(\frac{mv_F^2}{0.1eV}\Big)\cdot\Big(\frac{B}{40mT}\Big)\cdot\Big(\frac{\frac{1}{2}\cdot 10^{12}cm^{-2}}{n}\Big)^{\frac{3}{2}}\cdot 10^{-4}
\end{align}
for Dirac materials and
\begin{align}
\approx 4.8\cdot\Big(\frac{mv_F^2}{0.1eV}\Big)\cdot\Big(\frac{B}{40mT}\Big)\cdot\Big(\frac{\frac{1}{2}\cdot 10^{12}cm^{-2}}{n}\Big)^2\cdot 10^{-5}
\end{align}
for biased Bernal bilayer graphene. Semenoff energy gaps $mv_F^2$ on the order of $30$meV for graphene-hBN superlattices \cite{jung2017moire} and $1.5$eV for group-VI TMD's \cite{xiao2012coupled} lead to $|\gamma |\approx 3\cdot 10^{-3}$ in the former and $|\gamma |\approx 1.4\cdot 10^{-1}$ in the latter case. For biased Bernal bilayer graphene we estimate $|\tilde{\gamma}|=3.2\cdot 10^{-4}$ for $mv_F^2=0.1eV$. Notice that with $v_Fk_F(n)\approx 0.08eV$ we find $v_Fk_F\lesssim mv_F^2\lesssim\gamma_1$ as required for the validity of the chiral fermion model. Additional suppressions are expected due to only partial valley polarization and exponential decay of valley polarization due to intervalley scattering at the channel boundary. A realistic estimate for group-VI TMD's without these suppressions lets us expect a valley Hall viscosity signal to be around one to two orders of magnitude smaller than that of Hall viscosity which stands out as compared to graphene-based materials. This favors group-VI TMD's over graphene-hBN heterostructures and probably as well biased bilayer graphene as a platform for future experimental efforts to detect valley Hall viscosity. This requires to realize hydrodynamic electron flow in monolayer TMD's or appropriate TMD-based heterostructures and a corresponding Hall viscosity measurement first, though.

\section{Conclusions}
In this work we discussed the valley-resolved viscous Hall effect for (2+1)D Dirac materials. We motivated the validity of a finite valley Hall viscosity and well-defined valley-resolved Hoyos-Son formula for these materials following an earlier proposal for the valley-summed case \cite{sherafati2016hall,sherafati2019hall} by choosing a natural regularization and renormalization prescription.\par
Since early valley Hall conductivity theory and experiments focused on graphene-hBN heterostructures, group-VI TMD's and biased bilayer graphene, we chose these materials as explicit platforms on which valley-resolved Hall viscosity measurements might be performed in near future experiments. We provided estimates of expected signal strengths of valley Hall viscous transport within hydrodynamic electron flow. Our findings single out group-VI TMD based systems as the most promising platforms to detect valley-resolved Hall viscosity, even though TMD-based valley-summed Hall viscosity measurements analogous to \cite{berdyugin2019measuring,kim2025viscous} seem to be lacking as of now. Experimental detection could ultimately confirm our regularization prescription for a finitely renormalized valley Hall viscosity.\par
Recently, charge and valley hydrodynamics of gapped graphene in the quantum Hall regime has been investigated featuring a transport coefficient closely related to (valley) Hall viscosity, namely the stress response to inhomogeneous electric fields which is also fundamentally linked to nonlocal Hall conductivity \cite{shu2025charge}.  An interesting research avenue opened up by our work is the investigation of valley thermal Hall transport. We recently considered the latter for phonons in graphene geometrically coupled to electrons in the presence of a background magnetic field \cite{selch2026intrinsic}.


\appendix

\section*{Appendix}

\section{Reference appendix}
\label{appendixa}

We refer here to calculations of our previous work \cite{selch2026emergent} relevant for the main text and to be found there within the supplementary material appendices. In appendix A therein we discuss spinorial gauge transformations which allow to represent the effective Dirac action for Dirac materials in such a way that the gap enters in the form of a valley-odd mass. In appendix B1 we discuss the eigenproblem of massive Dirac fermions in a magnetic field. Appendix C discusses the proper non-relativistic limit of the eigenproblem keeping track of individual valley contributions. Appendix E provides details on the definitions of Hall conductivity and Hall viscosity relative to the electric current and the stress tensor. Details on Wigner-Weyl calculus, its application to transport and linear response theory may be found in appendices F, G and H, respectively. Finally appendix K exhibits the explicit calculations of $\mathcal{N}_{\sigma}$ and $\mathcal{N}_{\eta}$.\par

Details on our employed regularization prescription for valley-resolved Hall viscosity, its alternative extraction from the (finite frequency) relativistic Hoyos-Son formula following \cite{sherafati2019hall}, the eigenproblem of biased Bernal bilayer graphene in the low energy chiral fermion approximation as well as details on the corresponding (valley) Hall viscosity are presented in subsequent appendices.

\section{``Minimally relativistic'' regularization prescription for valley Hall viscosities}
\label{appendixb}

We discuss our ``minimally relativistic'' regularization prescription now directly relative to the work of Sherafati and Vignale \cite{sherafati2019hall}. The Green function representation of the (valley) Hall viscosity leads to a sum over a 2-tuple of states of which one is occupied, while the other one is unoccupied \cite{selch2026emergent}. This is familiar from the Kubo formalism to which our formalism is equivalent. A comparison of the sum for the Dirac theory and its non-relativistic limit yields as the main distinction an additional sum over particle-hole partner states in the former case plus sums involving zero Landau level states. Since the Dirac spectrum is not bounded from below in contrast to the non-relativistic theory, the possibility to sum over a particle-hole conjugate state allows to find pairs of occupied and unoccupied states which imply the sum to be infinite such that occupied states arbitrarily deep in the Dirac sea contribute to the sum representing individual valley Hall viscosities. We subtracted the resulting divergence in a natural way as explained in the main text to obtain the renormalized valley-selective Hall viscosities
\begin{align}
\eta^{K,K^\prime}_{H,ren}=\eta^{K,K^{\prime}}_{H,bare}-\eta^{K,K^{\prime}}_{H,vac}\,\,\,\,(\eta^{K^{\prime}}_{H,vac}=-\eta^{K}_{H,vac})
\end{align}
with bare and vacuum contributions $\eta^{K,K^{\prime}}_{H,bare}$ and $\eta^{K,K^{\prime}}_{H,vac}$, respectively. In the formulation of \cite{sherafati2019hall} the zero frequency valley-resolved Hall viscosities are represented by two subsums comprising functions $F_1$ and $F_2$. This separation corresponds exactly to our prescription according to the schematic form $\eta^{K,K^\prime}_{H,ren}\propto\sum_kF_1(k,\gamma )$, $\eta^{K,K^{\prime}}_{H,bare}\propto\sum_k(F_1(k,\gamma )+F_2(k,\gamma ))$ and $\eta^{K,K^{\prime}}_{H,vac}\propto\sum_kF_2(k,\gamma )$. The sum over $F_2$ is infinite and divergent. We thus drop the sum over $F_2$ completely. In particular this choice sets the value of the individual valley Hall viscosities to zero, if the chemical potential $\mu$ fulfills $-|m|<\mu <|m|$.
The prescription employed in this work is ``minimally relativistic'' in the following senses: (i) The sum over terms may be obtained from the non-relativistic theory, consisting of two spin degenerate, independent electron gases with relative Landau level occupation of one due to valley Zeeman splitting, by supplementing the additional sum over particle-hole partner states only for those occupied states appearing in the finite sum plus zero Landau level contributions. (ii) The sum comprising $F_1$ is finite and nonzero in the non-relativistic limit $\gamma\to\infty$, whereas that comprising $F_2$ vanishes if the limit is performed before the sum. The sum over $F_1$ involves only terms in a two Landau level neighborhood of the chemical potential. It produces both the expected result for the (valley-summed) Hall viscosity and a finite valley Hall viscosity which, as explained in the main text, is in conformity with the relativistic generalization of the Hoyos-Son formula applied to individual valleys. The latter property confirms that our prescription is compatible with nonlocal Hall conductivity in a natural way. In the case of a Galilean invariant theory, the relation between Hall viscosity and nonlocal Hall conductivity has been expressed rigorously via effective field theory methods and Ward identities, while their relation in the Lorentz invariant case has been suggested empirically only as explained in the main text. Finally note that Sherafati and Vignale obtained a finite, nonlocal, valley-resolved Hall conductivity but a divergent valley-resolved Hall viscosity. Their first nonlocal correction proportional to the squared momentum comprises a piece linear in Dirac mass which is valley-odd. This finding as well as experimental accessibility of the valley Hall viscosity transport coefficient suggest the existence of a finite transport valley Hall viscosity and the necessity of identifying a unique regularization prescription.

\section{Valley-resolved Hall viscosity at finite frequencies and the Hoyos-Son formula}
\label{appendixc}

We compare our valley-resolved Hall viscosity for Dirac materials with the local and nonlocal valley Hall conductivities of massive Dirac fermions as computed in \cite{sherafati2019hall} employing the relativistic Hoyos-Son formula discussed by the same authors. Following their notation we obtain
\begin{align}
\mathcal{N}_{\eta}^{\zeta}=\frac{1}{2}(G(p,s(\zeta )\gamma ,0)+G(p+1,s(\zeta )\gamma ,0)).
\end{align}
This result finally motivates an expression for finite frequency single valley Hall viscosities
\begin{align}
&\eta_H^{\zeta}=\frac{1}{2}(G(p,s(\zeta )\gamma ,\Omega )+G(p+1,s(\zeta )\gamma ,\Omega ))\frac{B}{2\pi}
\end{align}
with $\Omega =\frac{\omega}{\omega_0}$, $\omega_0=\frac{\sqrt{2}v_F}{l_B}$, magnetic length $l_B=\frac{1}{\sqrt{eB}}$ and
\begin{align}
\nonumber G(p,\gamma ,\Omega )=&\frac{1-p}{2}\Big(1+\frac{\gamma}{\sqrt{p+\gamma^2}}\Big)\\
&\cdot\frac{\Omega^2-(2\gamma^2-2\gamma\sqrt{p+\gamma^2}+2p-2)}{\Omega^4-4\Omega^2(\gamma^2+p-1)+4}.
\end{align}
The valley-resolved local Hall conductivities take the form \cite{sherafati2019hall}
\begin{align}
\sigma_H^{\zeta}=\frac{e^2}{2h}F(p,s(\zeta )\gamma ,0)
\end{align}
with
\begin{align}
\nonumber F(p,\gamma ,\Omega )=&-\Big(1+\frac{\gamma}{\sqrt{p+\gamma^2}}\Big)\\
&\cdot\frac{\Omega^2-(2\gamma^2-2\gamma \sqrt{p+\gamma^2}+2p-1)}{\Omega^4-2\Omega^2(2\gamma^2+2p-1)+1}.
\end{align}
The valley-resolved nonlocal Hall conductivity up to order $k^2$ has been found to be \cite{sherafati2019hall}
\begin{align}
\nonumber \sigma_H^{\zeta}(k,p,\gamma ,\Omega)=&\frac{e^2}{2h}\Big[F(p,s(\zeta )\gamma ,\Omega )\\
\nonumber &+(kl_B)^2\Big(G(p,s(\zeta )\gamma ,\Omega )\\
\nonumber &+G(p+1,s(\zeta )\gamma ,\Omega )\\
&-(p-\frac{1}{2})F(p,s(\zeta )\gamma ,\Omega ))\Big)\Big].
\end{align}
It can be seen that switching valleys $K\leftrightarrow K^{\prime}$ may be captured by a sign flip of $\gamma$. The extension of valley-selective Hall viscosity to finite frequencies is a new result predicted from the findings in \cite{sherafati2019hall}. Our results suggest a relativistically generalized Hoyos-Son formula as advertised in \cite{sherafati2016hall,sherafati2019hall} with additional valley resolution at finite frequencies0+ to be valid.

\section{The energy eigenvalues and eigenstates of biased Bernal bilayer graphene in the low energy regime in a magnetic field}
\label{appendixd}

The energy eigenvalues and eigenstates of biased Bernal bilayer graphene in a magnetic field may be obtained as follows. In addition to sublattice pseudospin, spin and valley degrees of freedom a layer index is introduced. The two dimensional layer space comprises two copies of monolayer graphene coupled by interlayer tunneling according to Bernal layer stacking. We consider the simplified case where only the largest interlayer coupling parameter $\gamma_1\approx 0.38$eV is retained. The so obtained description of bilayer graphene constitutes an inversion symmetric model. In order to produce a nonzero valley Hall viscosity inversion symmetry needs to be broken. In the case of monolayer graphene inversion symmetry was broken via a Semenoff mass gap. A convenient and controllable way of breaking inversion symmetry in bilayer graphene may be achieved by application of a bias voltage between the two layers. For small interlayer bias induced band gap $2\Delta =2mv_F^2\ll \gamma_1$ and small momenta $v_F|k|\ll 2\Delta$, the low energy action of bilayer graphene in flat space may be approximated by $H_{\zeta}=\bold{d}^{\zeta}\cdot\boldsymbol{\sigma}$ with
\begin{align}
\bold{d}^{\zeta}=\Big(-\frac{v_F^2}{\gamma_1}(D_x^2-D_y^2),-s(\zeta ) \frac{v_F^2}{\gamma_1}\{D_x,D_y\},\Delta \Big)^T.
\label{bilayerhamiltonian}
\end{align}
The Pauli matrices $\boldsymbol{\sigma}$ represent a two dimensional projection of the sublattice pseudospin and layer spaces into the space of low energy degrees of freedom which is a hybridization of the former spaces. For the K-valley $s(\zeta )=+1$ and for the K$^{\prime}$-valley $s(\zeta )=-1$. The covariant derivative $\bold{D}$ comprises a constant background magnetic field $B$ which we describe on the gauge field level in Landau gauge by $A_x=0$ and $A_y=Bx$. In this gauge $k_y$ is a good quantum number. The curly brackets denote an anti-commutator. We will subsequently suppress the Fermi velocity and set $v_F=1$.\par
We tackle the eigenvalue problem for the Hamiltonian in Eq. (\ref{bilayerhamiltonian}) by introduction of a new dimensionless variable $\xi =\sqrt{B}(x-k_y/B)$ such that $iD_x=k_x=-i\sqrt{B}\frac{\partial}{\partial \xi}$ and $iD_y=k_y-A_y=-\sqrt{B}\xi$. In terms of the ladder operators $\hat{a}=\frac{1}{\sqrt{2}}(\xi +\frac{\partial}{\partial \xi})$ and $\hat{a}^{\dagger}=\frac{1}{\sqrt{2}}(\xi -\frac{\partial}{\partial \xi})$ the Hamiltonian may be expressed as follows
\begin{align}
&H_{\zeta =K}=\begin{pmatrix}
\Delta & -\frac{2B}{\gamma_1}(\hat{a}^{\dagger})^2\\
-\frac{2B}{\gamma_1}\hat{a}^2 & -\Delta
\end{pmatrix},\\
&H_{\zeta =K^{\prime}}=\begin{pmatrix}
\Delta & -\frac{2B}{\gamma_1}\hat{a}^2\\
-\frac{2B}{\gamma_1}(\hat{a}^{\dagger})^2 & -\Delta
\end{pmatrix}.
\end{align}
As eigenstates we choose the ansatz
\begin{align}
\nonumber s(K) =+1:\,\,&\Psi_{\pm n}=(d_{\pm n}\phi_{n},c_{\pm n}\phi_{n-2}),\,\,n\geq 2,\\
&\Psi_1=(\phi_1,0),\,\,\Psi_0=(\phi_0,0)\\
\nonumber s(K^{\prime})=-1:\,\,&\Psi_{\pm n}=(c_{\pm n}\phi_{n-2},d_{\pm n}\phi_n),\,\,n\geq 2,\\
&\Psi_1=(0,\phi_1),\,\,\Psi_0=(0,\phi_0)
\end{align}
such that $\hat{a}\phi_n=\sqrt{n}\phi_{n-1}$ and $\hat{a}^{\dagger}\phi_n=\sqrt{n+1}\phi_{n+1}$, respectively. Together with the normalization condition $1=c_{\pm n}^2+d_{\pm n}^2$ the solution for the eigenstates yields
\begin{align}
c_{\pm n}=\sqrt{\frac{E_{\pm n}-s(\zeta )\Delta}{2E_{\pm n}}},\,\,\,\,d_{\pm n}=\pm\sqrt{\frac{E_{\pm n}+s(\zeta )\Delta}{2E_{\pm n}}}
\end{align}
for $n\geq 2$ with corresponding eigenenergies
\begin{align}
E_{\pm n}=\pm\sqrt{\Big(\frac{2B}{\gamma_1}\Big)^2n(n-1)+\Delta^2},\,\,\,\,E_{0,1}=s(\zeta )\Delta .
\end{align}

\section{(Valley) Hall viscosity coefficients for bilayer graphene}
\label{appendixe}

The Hall viscosity formula in Eq. (\ref{topologicalhallvis}) of the main text relies critically on the validity of the relations
\begin{align}
\nonumber \frac{\partial Q_W}{\partial e_a^i}=&i\frac{\partial Q_W}{\partial k_j}e_j^a\star (D_W)_i\Big\rvert_{e^i_a=\delta^i_a}\\
=&i(D_W)_i\star \frac{\partial Q_W}{\partial k_j}e^a_j\Big\rvert_{e^i_a=\delta^i_a}.
\label{lorentzinvariancerelations}
\end{align}
These relations are valid for systems with Lorentz invariance. For Galilean invariant systems a formula identical to Eq. (\ref{topologicalhallvis}) has be obtained \cite{selch2026nonrenormalization}. While bilayer graphene is Lorentz invariant, its low energy description as presented in this work is neither Lorentz nor Galilean invariant. We therefore need to check the validity of Eq. (\ref{lorentzinvariancerelations}) which is straightforward. We obtain
\begin{align}
&\Big(\frac{\partial H_{\zeta}}{\partial e_x^i}-i\frac{\partial H_{\zeta}}{\partial k_j}e_j^x\star (D_W)_i\Big)\Big\rvert_{e^i_a=\delta^i_a}=i\frac{B}{\gamma_1}(\delta_i^y\sigma^x-\zeta \delta_i^x\sigma^y),\\
&\Big(\frac{\partial H_{\zeta}}{\partial e_y^i}-i\frac{\partial H_{\zeta}}{\partial k_j}e_j^y\star (D_W)_i\Big)\Big\rvert_{e^i_a=\delta^i_a}=i\frac{B}{\gamma_1}(\delta_i^x\sigma^x+\zeta \delta_i^y\sigma^y).
\end{align}
In order to perform the calculation we used
\begin{align}
(D_W)_x\star (D_W)_y-(D_W)_y\star (D_W)_x=iB.
\end{align}
This means that for bilayer graphene Eq. (\ref{topologicalhallvis}) does not faithfully represent the Hall viscosity coefficient but only its leading piece in powers of the filling factor. The low energy description is algebraically much less involved, though, which is why we employ it in the main text. This case allows the use of symbolic python as a tool to determine the (valley) Hall viscosity. Let the filling factor $p\geq 3$ be defined as in the main text. A calculation of the Hall viscosity coefficient for the valley-summed case per spin degree of freedom results in
\begin{align}
\nonumber \mathcal{N}_{\eta}^s(p)=&2(p-1)^2[4(p-1)^4+(p-1)^2-2]\\
&\cdot [4(p-1)^2-1]^{-2}.
\end{align}
In the limit of large filling factor we obtain
\begin{align}
\mathcal{N}_{\eta}^s(p)\overset{p\gg 1}{=}\frac{p^2}{2}
\end{align}
which coincides with previous calculations \cite{hsiao2021timereversal}. 
Let us define the dimensionless parameter $\tilde{\gamma}=\frac{m\gamma_1}{2B}$. The valley-difference Hall viscosity coefficient is found to be
\begin{align}
\nonumber\mathcal{N}_{\eta}^d=&-\frac{\tilde{\gamma}}{2}\Big[\frac{(p-2)(p-1)}{(2p-3)\sqrt{p(p-1)+\tilde{\gamma}^2}}\\
&+\frac{(p-1)p}{(2p-1)\sqrt{(p+1)p+\tilde{\gamma}^2}}\Big]
\label{valleydifferencebilayer}
\end{align}
per spin degree of freedom. A comparison of $\mathcal{N}_{\eta}^d$ for monolayer graphene vs. bilayer graphene strongly supports that the calculated valley(-difference) Hall viscosity coefficient of bilayer graphene in Eq. (\ref{valleydifferencebilayer}) is exact. This follows both from the structural similarity as well as the scaling with magnetic field at constant charge carrier number density at large filling factor ($\mathcal{N}_{\eta}^d\sim \frac{1}{B}$). More precisely let us define the functions $f^{ml}(p)=p$ and $f^{bl}(p)=p(p-1)$ for monolayer (ml) and bilayer (bl) graphene. The dispersion relations for mono- and bilayer graphene (the latter in the low energy chiral fermion approximation) are given by
\begin{align}
&E^{ml}_{+p}=\frac{mv_F^2}{\gamma}\sqrt{f^{ml}(p)+\gamma^2},\\
&E^{bl}_{+p}=\frac{mv_F^2}{\tilde{\gamma}}\sqrt{f^{bl}(p)+\tilde{\gamma}^2}.
\end{align}
The valley Hall viscosity coefficients may be expressed in the forms
\begin{align}
\nonumber\mathcal{N}^d_{\eta}=&\frac{\gamma g_s}{4}\Bigg[\frac{f^{ml}(p-1)}{(f^{ml})^{\prime}(p-1)\sqrt{f^{ml}(p)+\gamma^2}}\\
&+\frac{f^{ml}(p)}{(f^{ml})^{\prime}(p)\sqrt{f^{ml}(p+1)+\gamma^2}}\Bigg]
\end{align}
for monolayer and
\begin{align}
\nonumber\mathcal{N}^d_{\eta}=&-\frac{\tilde{\gamma}g_{s}}{2}\Bigg[\frac{f^{bl}(p-1)}{(f^{bl})^{\prime}(p-1)\sqrt{f^{bl}(p)+\tilde{\gamma}^2}}\\
&+\frac{f^{bl}(p)}{(f^{bl})^{\prime}(p)\sqrt{f^{bl}(p+1)+\tilde{\gamma}^2}}\Bigg]
\end{align}
for bilayer graphene, respectively, with degeneracy factor $g_s=2$ for spin. The presented parametrization of the valley Hall viscosities of monoloayer and bilayer graphene suggests a pattern which possibly continues for multilayer graphene beyond two graphene layers

\end{document}